\documentclass[11pt]{article}

\usepackage[a4paper,margin=27mm]{geometry}
\usepackage[T1]{fontenc}
\usepackage[utf8]{inputenc}
\usepackage{amsmath}
\usepackage{amssymb}
\usepackage{booktabs}
\usepackage{caption}
\usepackage{enumitem}
\usepackage{float}
\usepackage{graphicx}
\usepackage{microtype}
\usepackage[numbers,sort&compress]{natbib}
\usepackage{siunitx}
\usepackage{xurl}
\usepackage[hidelinks]{hyperref}

\graphicspath{{../figures/}{figures/}}
\setlist[itemize]{leftmargin=1.35em}
\setlist[enumerate]{leftmargin=1.65em}
\newcommand{\RtxRunDir}{rtx5090\_stability\_6h\_20260621T200259Z}
\newcommand{\RtxSnapshots}{358}
\newcommand{\RtxElapsedMinutes}{360.0}
\newcommand{\RtxSmCountMin}{170}
\newcommand{\RtxSmCountMax}{170}

\newcommand{\RtxDriftCycles}{-0.029}
\newcommand{\RtxCorrMin}{0.999729}
\newcommand{\RtxGpuUtilMedian}{100.0}
\newcommand{\RtxMemUtilMedian}{99.0}
\newcommand{\RtxVramGiBMedian}{26.51}
\newcommand{\RtxTempMedian}{62.0}
\newcommand{\RtxPowerMedian}{384.0}

\IfFileExists{alibi_results.tex}{

\newcommand{\AlibiColo}{SOF}

\newcommand{\AlibiAS}{8866}
\newcommand{\AlibiASName}{VIVACOM-AS - Vivacom Bulgaria EAD, BG}
\newcommand{\AlibiAnchorsTotal}{48}
\newcommand{\AlibiAnchorsReachable}{30}
\newcommand{\AlibiBindingRadiusKm}{3168}

\newcommand{\AlibiDecoyRejected}{9}
\newcommand{\AlibiDecoyTotal}{11}
\newcommand{\DecoyRows}{%
Sydney & 13191 & rejected \\
Buenos Aires & 8416 & rejected \\
Singapore & 6960 & rejected \\
Sao Paulo & 6756 & rejected \\
Tokyo & 6090 & rejected \\
San Francisco & 6051 & rejected \\
Johannesburg & 5483 & rejected \\
Mumbai & 3264 & rejected \\
New York & 3158 & rejected \\
Reykjavik & 0 & accepted \\
Sofia (claimed) & 0 & accepted \\}

\newcommand{\CrossDieRows}{%
box1: RTX 5090 & 170 & 31 & 96 & 2.47 & \texttt{0d6493d6862b} \\
box2: RTX PRO 6000 & 188 & 95 & 128 & 2.62 & \texttt{d6397cf59fa3} \\
box3: B200 & 148 & 178 & 126 & 1.97 & \texttt{d09bfa963968} \\
box4: V100 & 80 & 32 & 6 & 1.53 & \texttt{b35848e4f865} \\}
}{}
\IfFileExists{crossdie_results.tex}{
\newcommand{\CrossBoxOneSM}{170}
\newcommand{\CrossBoxTwoSM}{188}

\newcommand{\CrossWithinCorrOneMin}{0.99981}
\newcommand{\CrossWithinCorrTwoMin}{0.99983}
\newcommand{\CrossSigCorr}{0.6206}

\newcommand{\CrossMeanLatTwo}{350.5}
\newcommand{\CrossLatDprime}{0.53}
\newcommand{\CrossShapeAcc}{100.0}
\newcommand{\CrossNSamples}{358}
}{}
\IfFileExists{noc_results.tex}{
\newcommand{\NocPeakGBps}{1637}
\newcommand{\NocKneeWarps}{3008}

\newcommand{\NocLowLoadGBps}{11}
\newcommand{\NocLowLoadWarps}{8}
\newcommand{\NocLowLoadLatNs}{11.6}
\newcommand{\NocSatLatNs}{0.08}
\newcommand{\NocLatHideFactor}{149}

}{}
\IfFileExists{nvhbi_results.tex}{
\newcommand{\NvhbiNSM}{148}
\newcommand{\NvhbiRegions}{64}
\newcommand{\NvhbiDieA}{74}
\newcommand{\NvhbiDieB}{74}
\newcommand{\NvhbiSameDie}{278}
\newcommand{\NvhbiCrossDie}{308}
\newcommand{\NvhbiPenaltyCyc}{30}
\newcommand{\NvhbiPenaltyNs}{15.5}
\newcommand{\NvhbiPenaltyPct}{10.9}
\newcommand{\NvhbiAntiCorr}{-0.76}
}{}
\IfFileExists{b200net_results.tex}{
\newcommand{\BnetMid}{182341274}
\newcommand{\BnetProbes}{169}
\newcommand{\BnetInliers}{168}

\newcommand{\BnetBindingKm}{79}
\newcommand{\BnetBindingMs}{0.79}
\newcommand{\BnetConsensusErrKm}{44}
\newcommand{\BnetDecoysRej}{11}
\newcommand{\BnetDecoysTotal}{11}
}{}
\IfFileExists{v100_results.tex}{

\newcommand{\VtenNetMid}{182343151}
\newcommand{\VtenNetProbes}{167}
\newcommand{\VtenNetBindingKm}{298}
\newcommand{\VtenNetErrKm}{270}
\newcommand{\VtenNetDecoysRej}{11}
\newcommand{\VtenNetDecoysTotal}{11}
}{}
\IfFileExists{stability_results.tex}{
\newcommand{\StabSnapshots}{358}
\newcommand{\StabHours}{6.0}
\newcommand{\StabCorrMin}{0.999798}

\newcommand{\StabNoiseFloorCyc}{0.09}
\newcommand{\StabNoiseFloorPct}{0.025}
\newcommand{\StabSigRatio}{342}
}{}
\IfFileExists{memtopo_results.tex}{
\newcommand{\MemTopoRows}{%
V100 & Volta & 80 & unified L2 & -- & -0.00 & -- \\
H200 & Hopper & 132 & two-way L2 split & 66/66 & -0.95 & 31 \\
B200 & Blackwell & 148 & two-die NV-HBI & 74/74 & -0.76 & 30 \\}
\newcommand{\MemVtenAnti}{-0.00}
\newcommand{\MemHtwoSplit}{66/66}
\newcommand{\MemHtwoAnti}{-0.95}
\newcommand{\MemHtwoPen}{31}
}{}

\title{\textbf{Unprivileged Topology Certificates for Cloud GPU Attestation}}

\author{Faruk Alpay$^{1*}$ \quad Taylan Alpay$^2$\\[0.35em]
\small $^1$\,Department of Computer Engineering, Bahçeşehir University, Istanbul, Türkiye\\
\small $^2$\,Department of Aerospace, University of Turkish Aeronautical Association, Ankara, Türkiye\\
\small \texttt{faruk.alpay@bahcesehir.edu.tr} \quad \texttt{s220112602@stu.thk.edu.tr}\\[0.25em]
\small *Correspondence: \texttt{alpay@lightcap.ai}}
\date{}

\begin{document}
\maketitle

\begin{abstract}
Cloud GPU tenants are asked to trust provider claims that are hard to inspect
from inside a rented job.  The accelerator should be the same physical machine
over time, it should match the billed hardware class, and it should run at the
advertised site.  Vendor attestation can answer related questions on
confidential-computing parts, but many rented accelerators expose no such path to
an unprivileged tenant.  We present a software-only attestation primitive for
that setting.  A CUDA probe measures an SM-by-memory-region
latency matrix using physical
\texttt{\%smid} labels and dependent global loads.  A streaming reducer commits
the resulting sufficient statistics, configuration, code hashes, and network
evidence into a small SHA-256 certificate that a verifier checks without a GPU.
The certificate supports three claims.  First, the per-SM latency map is a stable
physical fingerprint.  Over a six-hour full-load RTX 5090 run its median temporal
jitter is \StabNoiseFloorCyc{} cycles, while shape-only leave-one-out
classification separates distinct Blackwell dies with \CrossShapeAcc\% accuracy.
Second, cache-bypassing sweeps over HBM recover hardware-class topology across
generations, including a unified Volta V100 memory domain, a two-way Hopper H200 L2 split,
and a Blackwell B200 two-die NV-HBI package whose \NvhbiDieA/\NvhbiDieB{} SM
partition carries a \NvhbiPenaltyCyc-cycle (\NvhbiPenaltyNs~ns) cross-die penalty.
Third, public network landmarks bind the same certificate to a coarse location.
\BnetProbes{} RIPE Atlas probes place the B200 within \BnetConsensusErrKm{}~km of
its claimed datacentre and reject all \BnetDecoysTotal{} decoy sites.  Together,
these measurements check cloud-GPU identity, class, and coarse location without
privileged access or a vendor key.
\end{abstract}

\section{Introduction}

Renting a GPU in the cloud means running on a machine the tenant cannot inspect.
The provider reports a model name, a region, and a price.  The tenant must decide
whether the physical die is the advertised one, whether the job moved to another
accelerator, and whether the machine is in the jurisdiction named by the
contract.  These are attestation questions, but the usual answer relies on a
trusted hardware root and vendor signing keys
\cite{rfc9334,nvidia-gpu-cc-2025}.  That path is not available to every rented
GPU, especially consumer and workstation parts exposed through commodity
marketplaces.

We study the weaker but widely available case in which an unprivileged tenant
can launch CUDA kernels and measure the network path to the host, but cannot
read privileged counters, firmware state, serial numbers, or vendor attestation
reports.  The adversary is the provider or reseller
operating the host.  It may substitute a different accelerator, move a job after
enrolment, misstate the datacentre location, or replay evidence produced by
another machine.  We do not try to prove firmware integrity or protect secrets
in use.  The target is narrower and directly checkable.  We bind a rented GPU
instance to a physical timing fingerprint, a hardware-class topology, and a
coarse network location.

The main contribution is a \emph{topology certificate}.  The tenant measures a
latency matrix over physical SMs and memory regions, reduces the raw stream to
bounded sufficient statistics, and hashes those statistics together with the
probe source, configuration, device metadata, and network-location evidence.  A
verifier recomputes the decision predicates from the summary and checks the
hashes.  Verification needs neither a GPU nor the full raw trace.  The
certificate is a security-systems object built from hardware timing measurements;
confidential-computing attestation remains the stronger primitive when the
hardware exposes it.

We evaluate the primitive on five rented GPUs, a Volta V100, a Hopper H200, and
three Blackwell parts (RTX 5090, RTX PRO 6000, and B200).  The measurements are
organized around three attestation claims.

\begin{enumerate}
  \item \textbf{Physical identity.}  A per-SM latency map remains stable under a
        six-hour full-load RTX 5090 run, with median temporal jitter
        \StabNoiseFloorCyc{} cycles and minimum map correlation \StabCorrMin{}.
        Across distinct Blackwell dies, shape-only leave-one-out classification
        reaches \CrossShapeAcc\%, giving a vendor-free substitution check.
  \item \textbf{Hardware class.}  Cache-bypassing HBM sweeps recover topology
        signatures that match architectural class, including unified Volta memory,
        partitioned Hopper L2, and the B200's two-die NV-HBI package.  The B200
        split supplies the strongest example of the class-attestation signal.
  \item \textbf{Coarse location.}  Public network landmarks give
        distance-bounding evidence for the same enrolled instance.  In the B200
        run, \BnetProbes{} RIPE Atlas probes place the host within
        \BnetConsensusErrKm{}~km of the claimed datacentre and reject all
        \BnetDecoysTotal{} decoy locations.
\end{enumerate}

The remaining measurements support these claims.  The certificate layer reduces
long runs to bounded verification summaries, and the source package includes the
compressed raw data archive.  The on-chip-network sweep explains why the
fingerprint is stable under load and records a contention surface relevant to
multi-tenant security.

\section{Related Work}

\paragraph{Trusted hardware attestation.}
Remote attestation establishes machine properties to a relying party
\cite{rfc9334}.  NVIDIA confidential computing extends this to the GPU with a
hardware-fused device identity and signed measurements of firmware and
configuration \cite{nvidia-gpu-cc-2025}.  These mechanisms are strong but require
a trusted execution environment on data-centre parts and trust in the vendor key
hierarchy; they neither apply to an unprivileged tenant on a consumer Blackwell
card nor speak to where the machine physically is.  Our primitive covers a
different operating point, with no special hardware, no vendor key, and an added
location plane.

\paragraph{Verifiable computation and telemetry.}
Decentralised-compute and telemetry systems seek to verify that remote work or
measurements are genuine.  Trustless GPU validation checks computational
\emph{correctness} through probabilistic recomputation and profiling, explicitly
not the physical identity of the executing die \cite{trustless-gpu-2025};
verifiable network telemetry proves properties of flow records with
zero-knowledge proofs \cite{verifiable-telemetry-hotnets25}.  Neither binds the
result to a specific physical accelerator at a specific network location, which
is the gap this paper addresses.

\paragraph{GPU microbenchmarking and side channels.}
The probe builds on memory-hierarchy microbenchmarking
\cite{mei2017dissecting,jia2018volta,jin2024noc,jarmusch2025blackwell} and on GPU
side-channel work that already shows per-device timing distinguishability
\cite{naghibijouybari2018rendered,dutta2023spy,zhang2025nvbleed}.  Here those
timing effects serve as an attestation identity.

\paragraph{Network geolocation.}
Constraint- and delay-based IP geolocation infers position from round-trip times
to vantage points of known location \cite{ripe-ipmap-2020}, and is known to be
evadable when used adversarially as a forensic tool \cite{gill2010dudewheres}.
We use it in the easier, falsification direction---rejecting inconsistent claimed
locations---and combine it with edge-colocation and origin-AS evidence, echoing
recent ``trust but verify'' treatments of operator-reported geolocation
\cite{operator-geo-2024}.

\section{Attestation Model}

Let $R$ be the raw stream produced by a remote GPU run, consisting of tuples such
as
\[
  (\mathrm{snapshot}, t, \mathrm{smid}, \mathrm{probe}, L, m),
\]
where $L$ is the measured latency and $m$ is run metadata.  The reducer maps $R$
to a bounded summary $S=f(R)$ containing counts, means, variances, extrema, and
the run-level metadata needed by the claim.  The certificate is
\[
  C = H(\mathrm{source}) \parallel H(\mathrm{config}) \parallel H(S)
      \parallel H(R_\mathrm{archive}) \parallel M,
\]
where $M$ records the decision rule, device metadata, and external archive
policy.  Verification checks the hashes and recomputes the claim from $S$; it
does not require access to a GPU or to the full raw stream.

The minimal verification payload scales with the claim surface.  Experiment
duration affects the raw stream:
\[
  |S| = O(|\mathrm{SM}|\,|\mathrm{probe}|\,|\mathrm{mode}|),
  \qquad
  |R| = O(|\mathrm{SM}|\,|\mathrm{probe}|\,|\mathrm{repetition}|\,T).
\]
The raw stream remains scientifically useful and is archived with the artifact;
the verifier's decision rule operates on the bounded summary $S$.
If $g$ is the decision rule for the paper's claim, the verifier computes
$g(S)$ and checks
\[
  V(C,S)=1
  \Longleftrightarrow
  H(S)=C_S,\; H(\mathrm{source})=C_Q,\; H(\mathrm{config})=C_K .
\]
The certificate records enough to check the claim and its provenance without
requiring reviewers to re-run the remote GPU.

\begin{figure}[H]
\centering
\includegraphics[width=\linewidth]{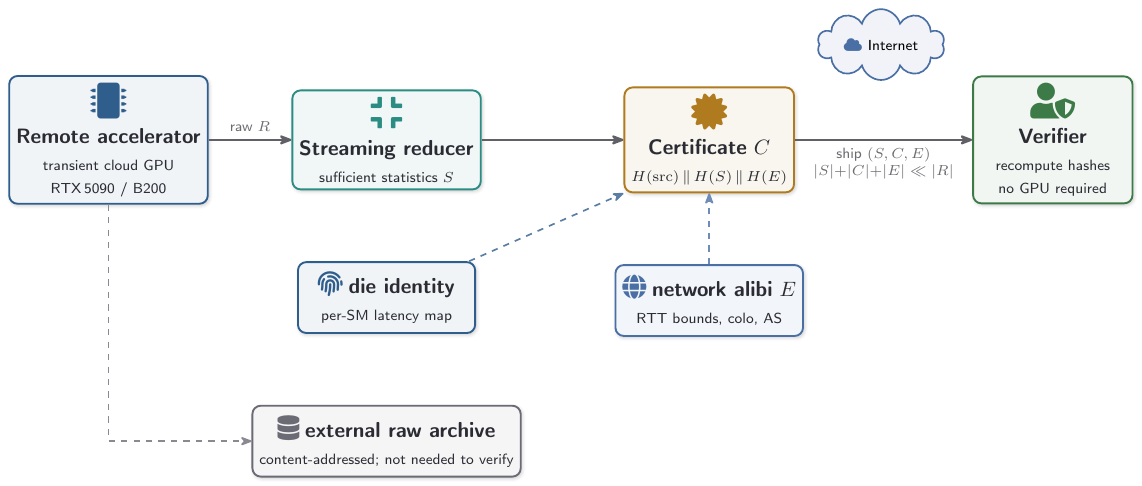}
\caption{The attestation pipeline.  The remote GPU produces raw timing rows; the
artifact ships bounded summaries and certificates to verifiers, and the arXiv
package carries the compressed raw data archive when it fits the source budget.}
\label{fig:pipeline}
\end{figure}

\section{GPU Probe}

The probe follows standard GPU microbenchmarking practice
\cite{mei2017dissecting,jia2018volta,jin2024noc}.  Each block reads the physical
SM identifier through the PTX special register \texttt{\%smid}
\cite{cuda-ptx}.  Dynamic shared-memory use forces one resident block per SM.
Thread 0 of each block claims a global turn counter, waits until no other block
is in the timed region, warms a dependent pointer chain, and times a sequence of
\texttt{ld.global.cg.u32} loads.  The \texttt{.cg} path avoids private-L1 hits,
so repeated dependent access to a self-referencing cache line is an L2/global
cache path measurement.

For a memory-region sweep, the output is a matrix $L(s,r)$ over physical SM
$s$ and probe region $r$.  For stability runs, the output is a sequence of
fingerprint matrices $F_k(s,p)$ indexed by snapshot $k$.  The artifact records
both cycles and telemetry because clock sources can differ.  CUDA device
attributes, \texttt{nvidia-smi} telemetry, and \texttt{clock64()} are related
but provide distinct operational views.  The primary stability claim is stated
in cycles and correlations unless a calibrated conversion to time is explicitly
reported.

\section{Certificate Stability Under Load}

We run the certificate layer on a busy device to avoid relying on idle
microbenchmark conditions.  The six-hour RTX 5090 run allocates
approximately \RtxVramGiBMedian{} GiB of VRAM, keeps the GPU at high memory and
compute utilization, and records one topology-fingerprint snapshot per minute.
The table below is generated from the JSON summary emitted by that run.

\begin{table}[H]
\centering
\caption{RTX 5090 stability artifact generated from the run summary.}
\label{tab:rtx-status}
\begin{tabular}{lr}
\toprule
Quantity & Value \\
\midrule
Run directory & \texttt{\RtxRunDir} \\
Observed duration & \RtxElapsedMinutes{} min \\
Topology snapshots & \RtxSnapshots{} \\
SM count per snapshot & \RtxSmCountMin{}--\RtxSmCountMax{} \\
Median GPU utilization & \RtxGpuUtilMedian\% \\
Median memory utilization & \RtxMemUtilMedian\% \\
Median allocated VRAM & \RtxVramGiBMedian{} GiB \\
Median temperature & \RtxTempMedian{} C \\
Median power & \RtxPowerMedian{} W \\
Minimum correlation vs first map & \RtxCorrMin{} \\
Mean-latency drift & \RtxDriftCycles{} cycles \\
\bottomrule
\end{tabular}
\end{table}

The summary is also a quality-control check.  Missing SMs, a low map
correlation, or a monotone latency drift would flag a failed measurement.  In
this run all 170 SMs appear in the snapshot set.  Median GPU utilization is
\RtxGpuUtilMedian\%, median memory utilization is \RtxMemUtilMedian\%, and the
minimum map correlation remains \RtxCorrMin{} while mean latency changes by only
\RtxDriftCycles{} cycles.

The same run fixes the measurement's noise floor.  Over \StabHours{} hours and
\StabSnapshots{} snapshots the median per-SM temporal jitter is
\StabNoiseFloorCyc{} cycles, or \StabNoiseFloorPct\% of mean access latency, and
the full map remains correlated with the first snapshot at \StabCorrMin{}.  We
use this floor as the significance scale for later topology claims.  The B200
NV-HBI penalty of \NvhbiPenaltyCyc{} cycles is \StabSigRatio$\times$ larger than
the jitter, putting the die split well above measured drift.

\clearpage
\section{Cross-Die Fingerprint Attestation}

The fingerprint matrix $F(s,p)$ is the candidate hardware identity.  An
attestation identity needs two properties.  It must remain stable for one
physical die across time and operating conditions, and it must differ between
physical dies in a way a verifier can check.  We test both properties on two
physical Blackwell dies measured with the identical probe, an RTX 5090
(\CrossBoxOneSM{} SMs, box~1) and an RTX PRO 6000 (\CrossBoxTwoSM{} SMs, box~2).

Stability holds under adversarial load.  Box~2 was driven through allocations of
$80$, $90$, and $60$~GiB at full utilization; across the snapshots of a single
run the full $\mathrm{SM}\times\mathrm{probe}$ map correlates with the first
snapshot at no less than $\CrossWithinCorrTwoMin$; box~1 reaches
$\CrossWithinCorrOneMin$.  The device mean latency is invariant to the
allocation size, about $\CrossMeanLatTwo$ cycles whether $80$ or $90$~GiB are
resident.  These measurements make the fingerprint a die property, independent
of the resident working set.

Separability holds at the same time.  Table~\ref{tab:crossdie} lists the
device descriptors and the measured device signature.  The two dies differ in
manufacturer-fixed parameters (SM count, L2 capacity), and also in latency
\emph{geometry}.  The $32$-probe device signatures correlate at
only $\CrossSigCorr$, and the per-SM mean latencies are separated with effect
size $d'=\CrossLatDprime$.  To rule out that this is merely a global offset or an
SM-count artifact, we shape-normalise each per-SM probe vector (zero mean, unit
variance) and run a leave-one-out nearest-neighbour classifier over all
$\CrossNSamples$ per-SM vectors from both dies.  It assigns each vector to the
correct die with $\CrossShapeAcc\%$ accuracy.  A verifier holding an enrolled
fingerprint for box~1 thus rejects box~2 as a substitute, and vice versa,
without any trusted execution environment, vendor certificate, or privileged
access---using only an unprivileged timing kernel.

\begin{table}[H]
\centering
\caption{Two enrolled physical Blackwell dies.  The dies are separable both by
fixed descriptors and by measured latency geometry (device-signature correlation
$\CrossSigCorr$; shape-only per-SM classification $\CrossShapeAcc\%$).}
\label{tab:crossdie}
\begin{tabular}{lrrrrl}
\toprule
Device & SMs & VRAM (GiB) & L2 (MiB) & Clock (GHz) & UUID prefix \\
\midrule
\CrossDieRows
\bottomrule
\end{tabular}
\end{table}

Two physical instances of two different products do not, by themselves,
establish same-model separability; that stronger claim is supported by prior
same-model GPU fingerprinting at the die level
\cite{dutta2023spy,naghibijouybari2018rendered}.  Our measurement shows that the
topology fingerprint remains stable under sustained full-GPU, near-full-VRAM
load and can be checked by an unprivileged tenant on commodity Blackwell silicon.

\clearpage
\section{Attesting Hardware Class}
\label{sec:class}

The second claim is hardware class.  A provider that bills for one accelerator
and schedules another should have to reproduce the memory topology of the billed
part, not just its model string.  The probe measures an architecture-level
partition signature.  Because a self-referencing line can be
served from L2 with nearly uniform latency, the class probe reads the physical
home of each address.  It uses 64-bit addressing over a buffer filling most of
HBM, places \NvhbiRegions{} targets uniformly across the buffer, and reads each
target with cache-bypassing (\texttt{ld.global.cv}) dependent loads.  Treating
the resulting $\mathrm{SM}\times\mathrm{region}$ latency matrix as a weighted
graph and bipartitioning its SMs by the Fiedler vector exposes how the device's
memory system is partitioned.

The signature differs by architecture (Table~\ref{tab:memtopo}).  A Volta V100
has a unified L2 and shows no partition.  Its two SM groups are uncorrelated
($r=\MemVtenAnti$), as expected for a single memory domain.  A Hopper H200
exposes its two-way partitioned L2, with \MemHtwoSplit{} SMs, strongly
anti-correlated profiles ($r=\MemHtwoAnti$), and a \MemHtwoPen-cycle
cross-partition penalty.  A Blackwell B200 exposes a two-die package joined by
NV-HBI.  Its \NvhbiNSM{} SMs split \NvhbiDieA/\NvhbiDieB{}, the per-reticle
counts, with anti-correlation $r=\NvhbiAntiCorr$ and a $2\times2$ block of
same-die latency \NvhbiSameDie{} and cross-die latency \NvhbiCrossDie{} cycles.
The NV-HBI crossing penalty is \NvhbiPenaltyCyc{} cycles
(\NvhbiPenaltyNs~ns, \NvhbiPenaltyPct\%), which is
\StabSigRatio$\times$ the identity noise floor measured under load.

\begin{table}[H]
\centering
\caption{Memory-partition signature recovered by the cache-bypassing probe across
three GPU generations.  Each part exposes a different topology---unified,
partitioned L2, and a two-die package---so the signature an instance presents must
match the class it claims.}
\label{tab:memtopo}
\begin{tabular}{llrlrrr}
\toprule
Part & Generation & SMs & Memory topology & Split & $r$ & Penalty (cyc) \\
\midrule
\MemTopoRows
\bottomrule
\end{tabular}
\end{table}

The attestation point is comparative.  A unified-L2 Volta, a two-partition
Hopper, and a two-die Blackwell are mutually distinguishable from timing alone.
Within a class, the residual per-SM geometry feeds the physical-identity
fingerprint from the previous section.  The signatures here come from rental
instances of limited tenure.  A longer per-part campaign
(\texttt{b200\_nv\_hbi.json}) repeats the sweep under several seeds for a
bootstrap interval on the cut conductance, at about 2.5\,s per full-device
snapshot.

\clearpage
\section{Location Attestation}

The fingerprint identifies the die and the topology probe checks its class.  The
third claim is coarse location.  The verifier cannot trust the provider's region
string, but it can use the fact that signals obey the speed of light.  Alongside
the certificate it records a small location object
\[
  E = \{(u_i,k_i,o_i,\tau_i,h_i)\}_{i=1}^{n},
\]
where $u_i$ is a landmark, $k_i$ its role, $o_i$ a success bit, $\tau_i$ a
round-trip or fetch latency, and $h_i$ a response hash, and binds it by
\[
  C_\mathrm{loc}=H(\mathrm{host}\parallel \mathrm{created}\parallel E),
\]
so altering any landmark, response, or host context changes the location
commitment without touching the device statistic $S$.  The verifier uses $E$ to
compute the distance bound; the bulk timing stream $R$ is separate from this
minimal location object.

With timing, $E$ bounds location.  Cloudflare supplies edge path metadata,
M-Lab's ndt7 locate service
identifies nearby active-measurement servers, and RIPE Atlas exposes anchored
vantage points with published coordinates
\cite{cloudflare-speedtest,mlab-ndt7,ripe-atlas-measurements,ripe-ipmap-2020}.
For a vantage point $v$ at known coordinates with measured minimum round-trip
time $\rho_v$ to the instance, propagation cannot beat the speed of light, so the
instance must lie within a disk centred on $v$ of radius
\[
  r_v \;=\; \tfrac{1}{2}\,\rho_v\,\kappa c, \qquad \kappa \approx \tfrac{2}{3},
\]
where $\kappa c$ is a conservative effective fibre velocity.  The intersection
$\bigcap_v D(v,r_v)$ is a constraint region for the instance, and any claimed
location outside it is falsified.  Constraint-based geolocation
\cite{ripe-ipmap-2020,gill2010dudewheres} provides the certificate-bound
location \emph{alibi}.

We measured $\rho_v$ to \AlibiAnchorsReachable{} of \AlibiAnchorsTotal{} anchored
vantage points; the tightest (binding) disk has radius \AlibiBindingRadiusKm{}~km.
The edge colocation \AlibiColo{} and the origin autonomous system AS\AlibiAS{}
(\AlibiASName) place the instance independently, and both are consistent with the
RTT region.  The discriminating power of the region is shown in
Table~\ref{tab:decoy}.  Of \AlibiDecoyTotal{} globally distributed decoy
datacentre locations, \AlibiDecoyRejected{} are rejected as inconsistent with
the measured round-trip times, while the true site is accepted.

\begin{table}[H]
\centering
\caption{Falsification test.  Each candidate location is checked against the
intersection of speed-of-light disks; ``max violation'' is the largest distance
by which the candidate falls outside any single disk.  \AlibiDecoyRejected{} of
\AlibiDecoyTotal{} decoys are rejected; the claimed site is accepted.}
\label{tab:decoy}
\begin{tabular}{lrl}
\toprule
Candidate location & Max violation (km) & Verdict \\
\midrule
\DecoyRows
\bottomrule
\end{tabular}
\end{table}

The bound is intentionally conservative.  ICMP echoes are de-prioritised and can
follow indirect paths, so $\rho_v$ overestimates propagation and the disks are
larger than the true light-cone.  The resulting region localises the instance
only to continental scale and admits an occasional nearby false accept.  In this
run, the single high-latitude European false accept is resolved by the colocation
and autonomous-system evidence.  Tighter regions follow from measuring $\rho_v$
\emph{towards} the instance from many distributed probes with known positions.
For the B200 run we used RIPE Atlas \cite{ripe-atlas-measurements}.  Because the
instance does not answer ICMP, we scheduled a TCP traceroute to an open port from
a worldwide probe set (public measurement \BnetMid).  Of \BnetProbes{}
responding probes a single one reported a physically impossible
sub-millisecond round-trip from a distant registered location---an instance of
operator-reported geolocation error \cite{operator-geo-2024}---which a consensus
(maximum-inlier) fit discards.  The remaining \BnetInliers{} place the
instance within \BnetConsensusErrKm{}~km of its claimed datacentre, with a
binding disk of \BnetBindingKm{}~km from the nearest probe
(\BnetBindingMs~ms), and reject all \BnetDecoysRej{} of \BnetDecoysTotal{}
decoy locations.  The measurement identifier is recorded in the certificate, so a
third party re-fetches the same vantage points and recomputes the region without
trusting us.  The object the verifier checks is the path-context $E$, its hash
$C_\mathrm{net}$, and the decision predicate over the disks.

\section{Location Attestation as an Enforcement Primitive}

One application of location attestation is accelerator governance.  Recent policy
proposals argue that controlled AI chips should be able to prove that they run
only in approved jurisdictions, often by relying on on-chip roots of trust
\cite{cnas-governable-chips,iaps-location-verification,scher2024mechanisms}.  The
certificate studied here has weaker guarantees, but it applies to rented parts
that do not expose a confidential-computing attestation path.  It replaces a
cryptographic device identity with a physical fingerprint and combines that
fingerprint with a distance bound \cite{brands1993distance}.

Round-trip time is one-sided evidence.  For a landmark $v$ at known position, the
instance lies within $r_v=\tfrac12\rho_v\kappa c$, and an adversarial host can
only inflate $\rho_v$, not reduce it below true propagation.  A low RTT from a
landmark in region $A$ places the instance near $A$.
High RTT alone is not evidence of absence from a forbidden region, because the
host can add delay.  Enforcement deployments need nearby honest landmarks that
pin the feasible region.  Without them, the host can enlarge the disks or stop
responding.

Network measurement alone remains vulnerable to relays.  A controlled instance
could forward a challenge to a compliant proxy elsewhere and return the proxy's
timely reply \cite{brands1993distance}.  The fingerprint closes that gap.  The
reply is bound to the per-SM latency map of the responding die, and the proxy
does not reproduce the enrolled device's map.  The two layers cover different
failure modes.  Network timing supplies location information; the fingerprint
binds that timing to the enrolled die.

The measurements exercise both layers on parts with no confidential-computing
path.  The hardware-class signature (Table~\ref{tab:memtopo}) ties the controlled
part to its architecture, while the network proof ties the same certificate to a
place.  On the network side, \VtenNetProbes{} RIPE landmarks placed the V100
within \VtenNetErrKm{}~km of its declared site and rejected
\VtenNetDecoysRej{} of \VtenNetDecoysTotal{} decoys (public measurement
\VtenNetMid), while the B200 was pinned to \BnetConsensusErrKm{}~km.
Localization tightness follows landmark density: \BnetBindingKm{}~km near the
B200's dense region versus \VtenNetBindingKm{}~km where the nearest honest V100
landmark is farther away.

The guarantee is not tamper-proof.  A host with physical control can cut power,
refuse to run the probe, or avoid reachable services.  For cooperative hosts,
the certificate checks identity, class, and coarse location without relying on a
vendor key hierarchy, while keeping the failure modes explicit.

\clearpage
\section{Supporting On-Chip Network Measurement}
\label{sec:noc}

\begin{figure}[H]
\centering
\includegraphics[width=0.72\linewidth]{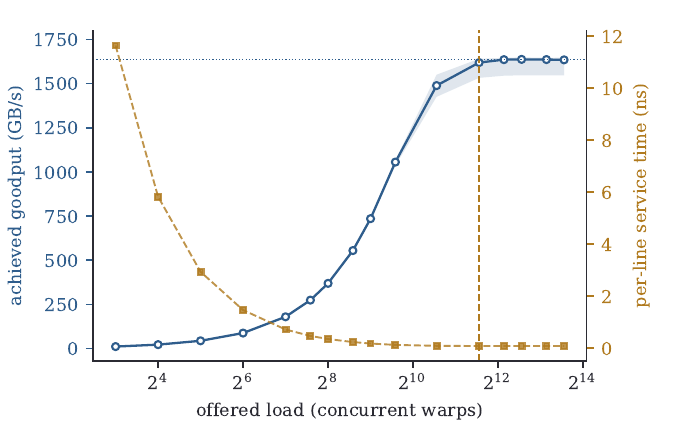}
\caption{The on-chip network as a congestion-controlled fabric (RTX PRO 6000,
85\,GiB resident).  Achieved goodput (left axis) rises with offered concurrency
and saturates at the dotted line, \NocPeakGBps{}~GB/s, past the knee marked by
the dashed vertical line near \NocKneeWarps{} concurrent warps; the effective
per-line service time (right axis) falls from \NocLowLoadLatNs{} to
\NocSatLatNs{}~ns as concurrency hides latency.}
\label{fig:noc}
\end{figure}

The per-SM latency structure used for identity is produced by the GPU's on-chip
interconnect.  SMs issue traffic to memory partitions through a routed fabric,
and placement on that fabric affects latency \cite{jin2024noc}.  To characterize
the substrate behind the fingerprint, we sweep the offered load, measured as the
number of warps concurrently injecting read traffic into a large resident buffer.  For each
load point we record delivered goodput and effective per-line service time
(Figure~\ref{fig:noc}).

The RTX PRO 6000 behaves like a latency-bound fabric at low concurrency and a
bandwidth-bound fabric after the knee.  At \NocLowLoadWarps{} concurrent warps it
delivers \NocLowLoadGBps{}~GB/s with an effective service time of
\NocLowLoadLatNs{}~ns per 128-byte line.  Goodput then rises almost linearly and
saturates at \NocPeakGBps{}~GB/s near \NocKneeWarps{} concurrent warps.  At
saturation the effective service time falls to \NocSatLatNs{}~ns, a
\NocLatHideFactor$\times$ reduction from the low-load regime.  The low-load
regime is where placement-dependent per-SM latencies are most visible, which is
why the same probe can produce a stable fingerprint.

The on-chip-network sweep supports the attestation claims.  The measured
structure belongs to the routed fabric, which explains why the fingerprint
survives heavy load.
The same fabric also appears in multi-tenant interference work, where co-located
tenants perturb timing and build contention channels
\cite{elvinger2025interference,zhang2025nvbleed}.  Repeating the sweep on other
generations puts the capacity on a scale.  Peak goodput is about 881\,GB/s on a
Volta V100 and 6986\,GB/s on a B200, against \NocPeakGBps{}~GB/s here, an
approximately $8\times$ span across the measured parts.  A multi-tenant
contention study is outside the scope of this certificate paper.

\clearpage
\section{Packaging}

The arXiv source bundle must compile cleanly and still carry enough evidence to
audit the measurements \cite{arxiv-format,arxiv-sizes}.  The local
\texttt{data/} directory is \num{76372647} bytes before compression, dominated
by CSV timing streams from the B200, H200, V100, and RTX 5090 affinity and
stability sweeps.  The package does not place those CSV streams directly in the
compile tree, and it avoids a nested tar or zip archive for the data payload,
since source checkers may account for such archives by their expanded contents.

The builder creates a fresh staging tree.  The root contains the files needed by
arXiv to compile the paper, including \texttt{main.tex}, generated result macros,
bibliography products, and figures.  The reproducibility material lives under
\texttt{anc/} \cite{arxiv-ancillary,arxiv-datasets}.  JSON summaries used for
quick inspection remain unpacked under \texttt{anc/data/}.  The complete
measurement directory is stored as \texttt{anc/data\_full.sqlite}.  Its
\texttt{files} table preserves the original \texttt{data/...} path, byte count,
SHA-256 digest, compression tag, compressed byte count, and payload for each of
the 35 evidence files.  Each payload is an LZMA-compressed copy of the original
file, so the store is reversible without dropping rows, reducing precision, or
sampling the timing streams.  Platform metadata is excluded.  A companion
\texttt{anc/data\_full\_manifest.tsv} records the same per-file accounting in a
plain text form.

The rest of \texttt{anc/} contains the CUDA kernels, Python reducers and
verifiers, experiment configurations, certificates, and a top-level
\texttt{artifact\_manifest.json} for the staged package itself.  The extraction
script \texttt{anc/scripts/extract\_data\_full.py} reconstructs the original
\texttt{data/} tree from the SQLite store and verifies the SHA-256 digest of
every restored file.  Before zipping, the builder rejects files that blur the
compile boundary, compiles the staged tree with the same
\texttt{pdflatex}/\texttt{bibtex} sequence used by arXiv, removes compile
by-products, and checks the final zip size.  The current submission zip is about
\num{16900000} bytes, and the staged file list totals about \num{17100000} bytes
while including the full raw evidence payload.

\section{Limitations}

The B200 measurement was taken on a rental instance with limited tenure.  It
establishes the two-die partition and the crossing penalty.  A bootstrap
confidence interval on the cut conductance awaits a longer multi-seed campaign.
The cross-die identity experiment separates two different Blackwell products.
Same-model die separation is left to prior GPU fingerprinting
\cite{dutta2023spy,naghibijouybari2018rendered}.  The network alibi localises to
metropolitan-to-continental scale, not to a rack, and
because the instances do not answer ICMP it uses TCP-reachable ports.  The
certificate proves consistency of the measured evidence; it does not prove
firmware integrity, prevent refusal to run, or replace vendor attestation when a
trusted TEE is available.

\section{Conclusion}

An unprivileged tenant can check more than a model name on an invoice.  Timing
measurements produce a stable per-SM identity, HBM sweeps expose a
hardware-class topology, and public landmarks bind the enrolled instance to a
coarse location.  Across five rented GPUs and three generations, these signals
fit into a small topology certificate that a verifier checks without a GPU.
Confidential-computing attestation remains stronger when available; this
vendor-free path covers substitution, class mismatch, and inconsistent location
claims on accelerators that expose no vendor attestation interface.

\bibliographystyle{unsrtnat}
\bibliography{refs}

\end{document}